\documentclass[10pt,conference]{IEEEtran}

\usepackage{fancyhdr}
\usepackage[normalem]{ulem}
\usepackage[sort,nocompress]{cite}
\usepackage[keeplastbox]{flushend}
\usepackage{acro}
\usepackage{booktabs}
\usepackage[enable]{easy-todo}
\usepackage{mathtools}
\usepackage{physics}
\usepackage{amsmath,amssymb,amsfonts}
\usepackage{algorithmic}
\usepackage{graphicx}
\usepackage{textcomp}
\usepackage{xcolor}
\usepackage{tikz}

\newcommand{\revision}[1]{{#1}}

\usepackage[bookmarks=true,breaklinks=true,letterpaper=true,colorlinks,citecolor=blue,linkcolor=blue,urlcolor=blue]{hyperref}

\DeclareAcronym{mu}{short=MU, long=machine units}
\DeclareAcronym{vqe}{short=VQE, long=variational quantum Eigensolver}
\DeclareAcronym{qpe}{short=QPE, long=quantum phase estimation}
\DeclareAcronym{qpu}{short=QPU, long=quantum processing unit}
\DeclareAcronym{cpu}{short=CPU, long=classical processing unit}
\DeclareAcronym{ibm}{short=IBM, long=International Business Machines}
\DeclareAcronym{dag}{short=DAG, long=directed acyclic graph}
\DeclareAcronym{artiq}{short=ARTIQ, long=advanced real-time infrastructure for quantum physics}
\DeclareAcronym{nisq}{short=NISQ, long=noisy intermediate-scale quantum}
\DeclareAcronym{dsl}{short=DSL, long=domain-specific language}
\DeclareAcronym{jit}{short=JIT, long=just-in-time}
\DeclareAcronym{rb}{short=RB, long=randomized benchmarking}
\DeclareAcronym{dds}{short=DDS, long=direct digital synthesizer}
\DeclareAcronym{dac}{short=DAC, long=digital-to-analog converter}
\DeclareAcronym{adc}{short=ADC, long=analog-to-digital converter}
\DeclareAcronym{awg}{short=AWG, long=arbitrary waveform generator}
\DeclareAcronym{fpga}{short=FPGA, long=field-programmable gate array}
\DeclareAcronym{yb171}{short=${}^{171}$Yb$^+$, long=Ytterbium 171}
\DeclareAcronym{pmt}{short=PMT, long=photomultiplier tube}
\DeclareAcronym{api}{short=API, long=application programming interface}
\DeclareAcronym{isa}{short=ISA, long=instruction set architecture}
\DeclareAcronym{rpc}{short=RPC, long=remote procedure call}
\DeclareAcronym{mw}{short=MW, long=microwave}
\DeclareAcronym{bb1}{short=BB1, long=broadband}
\DeclareAcronym{sk1}{short=SK1, long=Solovay-Kitaev}
\DeclareAcronym{spam}{short=SPAM, long=state preparation and measurement}
\DeclareAcronym{cw}{short=CW, long=continuous wave}
\DeclareAcronym{rtio}{short=RTIO, long=real-time I/O}
\DeclareAcronym{sqst}{short=SQST, long=single-qubit state tomography}
\DeclareAcronym{gst}{short=GST, long=gate set tomography}
\DeclareAcronym{1d}{short=1D, long=one-dimensional}
\DeclareAcronym{2d}{short=2D, long=two-dimensional}
\DeclareAcronym{ddb}{short=DDB, long=device database}
\DeclareAcronym{ast}{short=AST, long=abstract syntax tree}
\DeclareAcronym{rf}{short=RF, long=radio frequency}
\DeclareAcronym{mro}{short=MRO, long=method resolution order}
\DeclareAcronym{ir}{short=IR, long=intermediate representation}
\DeclareAcronym{vcd}{short=VCD, long=value change dump}
\DeclareAcronym{gpu}{short=GPU, long=graphics processing unit}
\DeclareAcronym{vliw}{short=VLIW, long=very long instruction word}
\DeclareAcronym{simd}{short=SIMD, long=single instruction multiple data}
\DeclareAcronym{hll}{short=HLL, long=high-level language}
\DeclareAcronym{ci}{short=CI, long=continuous integration}
\DeclareAcronym{hdl}{short=HDL, long=hardware description language}
\DeclareAcronym{mems}{short=MEMS, long=microelectromechanical systems}

\DeclareAcronym{dax}{short=DAX, long=Duke ARTIQ extensions}
\DeclareAcronym{max}{short=MAX, long=modular ARTIQ extensions}
\DeclareAcronym{staq}{short=STAQ, long=software-tailored architecture for quantum co-design}
\DeclareAcronym{rc}{short=RC, long=red chamber}
\DeclareAcronym{grape}{short=GRAPE, long=GRadient Pulse Engineering}
\DeclareAcronym{siqc}{short=SIQC, long=software integrated quantum computer}
\DeclareAcronym{aom}{short=AOM, long=acousto-optic modulator}
\DeclareAcronym{qci}{short=QCI, long=quantum-classical interface}
\DeclareAcronym{dlpc}{short=DLPC, long=device-level partial-compilation}

\newcommand*\circled[1]{\tikz[baseline=(char.base)]{
            \node[shape=circle,draw,inner sep=1pt] (char) {#1};}}

\title{One-Time Compilation of Device-Level Instructions for Quantum Subroutines} 
\author{
\IEEEauthorblockN{
Aniket~S.~Dalvi\IEEEauthorrefmark{1}\IEEEauthorrefmark{5},
Jacob~Whitlow\IEEEauthorrefmark{1},
Marissa~D'Onofrio\IEEEauthorrefmark{1},
Leon~Riesebos\IEEEauthorrefmark{1},
Tianyi~Chen\IEEEauthorrefmark{2},
Samuel~Phiri\IEEEauthorrefmark{1},\\
Kenneth~R.~Brown\IEEEauthorrefmark{1}\IEEEauthorrefmark{2}\IEEEauthorrefmark{3} and
Jonathan~M.~Baker\IEEEauthorrefmark{1}\IEEEauthorrefmark{4}
}

\IEEEauthorblockA{
\IEEEauthorrefmark{1}Duke Quantum Center and Department of Electrical and Computer Engineering, Duke University, Durham, NC
}
\IEEEauthorblockA{
\IEEEauthorrefmark{2}Department of Physics,
Duke University, Durham, NC
}
\IEEEauthorblockA{
\IEEEauthorrefmark{3}Department of Chemistry,
Duke University, Durham, NC
}
\IEEEauthorblockA{
\IEEEauthorrefmark{4}Department of Electrical and Computer Engineering,
University of Texas at Austin, Austin, TX
}
\IEEEauthorblockA{
\IEEEauthorrefmark{5}Email: aniketsudeep.dalvi@duke.edu
}
}

\begin{document}
\date{}
\maketitle

\thispagestyle{empty}

\begin{abstract}
A large class of problems in the current era of quantum devices involve interfacing between the quantum and classical system. These include calibration procedures, characterization routines, and variational algorithms. The control in these routines iteratively switches between the classical and the quantum computer. This results in the repeated compilation of the program that runs on the quantum system, scaling directly with the number of circuits and iterations. The repeated compilation results in a significant overhead throughout the routine. In practice, the total runtime of the program (classical compilation plus quantum execution) has an additional cost proportional to the circuit count. At practical scales, this can dominate the round-trip CPU-QPU time, between 5\% and 80\%, depending on the proportion of quantum execution time.

To avoid repeated device-level compilation, we identify that machine code can be parametrized corresponding to pulse/gate parameters which can be dynamically adjusted during execution. Therefore, we develop a device-level partial-compilation (DLPC) technique that reduces compilation overhead to nearly constant, by using cheap remote procedure calls (RPC) from the QPU control software to the CPU. We then demonstrate the performance speedup of this on optimal pulse calibration, system characterization using randomized benchmarking (RB), and variational algorithms. We execute this modified pipeline on real trapped-ion quantum computers and observe significant reductions in compilation time, as much as 2.7x speedup for small-scale VQE problems.
\end{abstract}

\section{Introduction}
\label{sec:introduction}

% Why is there gap - scale of QC operations is ns -> microseconds at worst, therefore when the scale of compilation is on this order it can eat up a large amoutn fot ime

% Let's think of a teaser image here showing the core message of the paper - i.e. the differentiation of different levels of compilation and the incorporation of partial compilation at the device level.

Current quantum computers are far from the requisite physical error rates and the qubits needed to support quantum error correction. 
% This is needed to execute algorithms like integer factorization~\cite{shor_factorisation}.
While hardware developers race toward this long-term goal, there is wide interest in finding near-term use cases for quantum computers and optimizing their software stack. One such class of problems is what we call \textit{\ac{qci}} problems, essentially programs, algorithms, or routines which repeatedly communicate with a quantum computer and make decisions about subsequent actions based on the outcomes. This class encompasses routines at both the algorithm/application layer as well as further down the hardware-software stack e.g. VQA~\cite{Peruzzo2014} and QAOA~\cite{hadfield2019quantum} which have similar circuit structures but vary parameters between iterations. This category also includes adaptive calibration or characterization routines~\cite{stace2022optimised, PhysRevA.77.012307, https://doi.org/10.48550/arxiv.1310.4492}.

\begin{figure}[h]
    \centering
    \includegraphics[width=\linewidth]{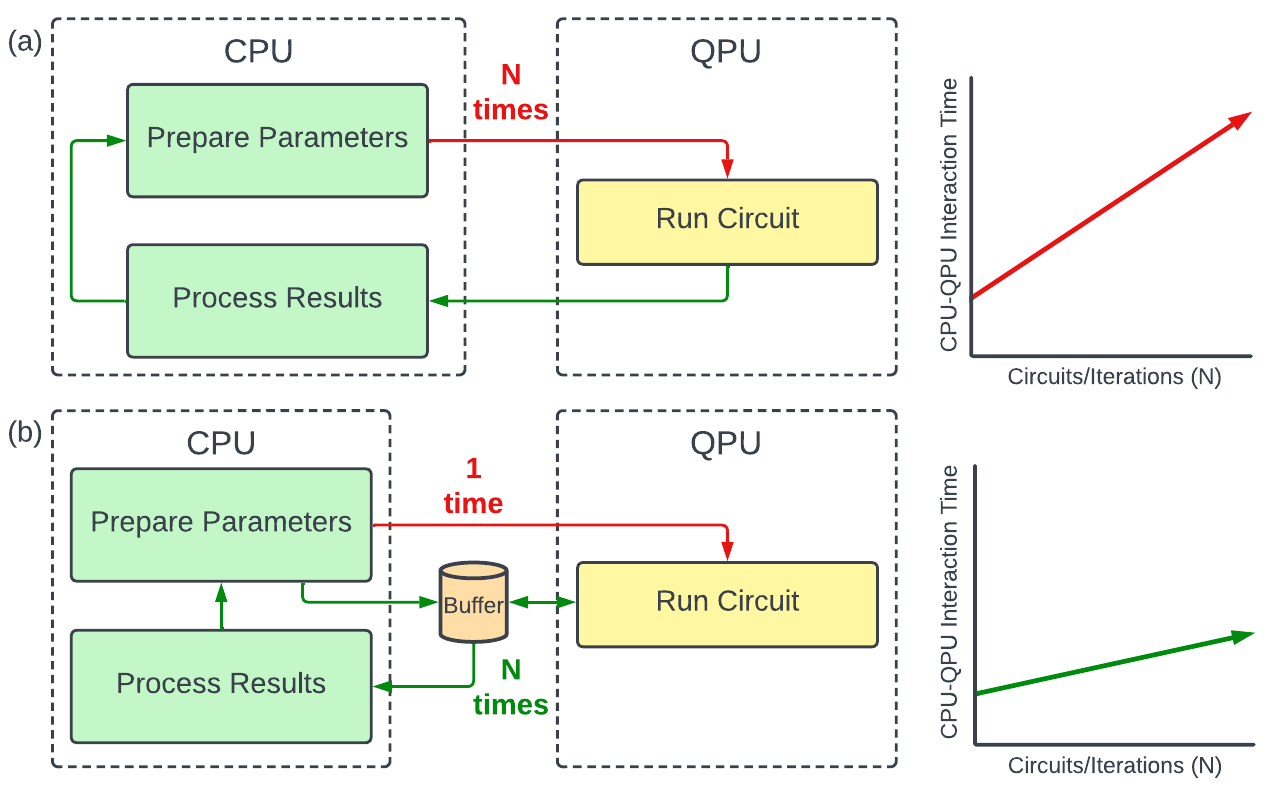}
    % \vspace{-1.5\baselineskip}
    \caption{The pipeline for execution of quantum programs on hardware involves a sequence of compilation steps - circuit level, pulse level, and device level. Device-level compilation involves the translation of circuit elements to control hardware primitives. For current applications and routines (a) the kernel is compiled every iteration and for every new set of circuits. Device compilation overheads scale as a function of the number of circuits. (b) Many circuits change in only small ways and have predictable effects on the underlying driver instructions.
    % , for example how long a pulse of a given frequency should play. 
    We propose partial compilation of the kernel which happens exactly once. In every iteration, we pass small amounts of information between the CPU and QPU and modify pulse execution appropriately.}
    % \vspace{-\baselineskip}
    \label{fig:intro}
\end{figure}

Common amongst each of these examples is a repeated back and forth between the quantum computer and the classical computer with varying amounts of intermediate \revision{classical} decision-making. The runtime of a quantum program involves compilation time and execution time. The execution time is often divided between classical execution on the \ac{cpu} and quantum execution on the \ac{qpu}, with varying ratios based on the type of program. Classical overheads are often ignored - we are willing to sacrifice large upfront classical compute times to optimize and compile input programs with the promise that quantum computers will provide sufficient speedup to accommodate this overhead. Recent work from IBM \cite{Kim2023}, however, showcases the implications of classical compilation overhead. Their total wall clock time for a program takes about 4 hours, with a device run time of just over 5 minutes.
Most quantum computing architectural studies have focused on optimization relatively high in the hardware-software stack, prioritizing circuit-level and pulse-level improvements. What has largely been ignored is the practical costs of preparing and executing these optimized programs on hardware. This is typically assumed to be very low relative to the execution time of the quantum program. 

\textit{Device-level compilation} refers to the conversion of the high-level intermediate representation of quantum gates or pulses to machine code runnable on the control hardware, similar to the translation from classical instruction set architecture (ISA) to executable binaries. In the current quantum computing stack, device-level compilation corresponds to the construction of a kernel (or a binary) that is run on the \ac{qpu} - the electronics that control the quantum hardware. 
% Execution typically starts with a high-level circuit which is compiled into optimized pulses. \revision{These} instructions are ready to be compiled to the \ac{qpu}. The pulses are converted to device driver instructions, which are eventually compiled to executable binaries to be run on the \ac{qpu}. This compilation of final optimized pulses to executable binaries encompasses device-level compilation.

For single-circuit applications, this amounts to only a single round of kernel compilation. For \ac{qci} problems, which require large numbers of circuits and/or iterations, repeated kernel compilation becomes substantial. This is illustrated in Fig.~\ref{fig:intro}(a).
Repeated sequential compilation of circuits, either by adding new subcircuit elements and/or modifying circuit parameters such as the angle of rotation, is expensive when the final solution for these algorithms requires hundreds or thousands of passes through the classical-quantum loop. Prior work has focused on reducing the compilation costs upstream from the actual execution of the circuit on hardware, for example, circuit optimization, gate decompositions, and scheduling. While minimizing this upfront compilation time is important, it often occurs only once during the execution of the entire program. Downstream optimizations, like pulse generation, might be repeated. The work in~\cite{gokhale_variational} proposes partial compilation techniques for variational algorithms to reduce the computational overhead of these repeated optimizations by only recompiling small sections of the circuit. 

High-level representations of programs also vary widely, and a shared structure is apparent only when these high-level programs are reduced to low-level hardware-specific representations, or \emph{device-level representations}. As a result, we focus on enabling optimal compilation through \ac{dlpc}. We propose a partial compilation technique, further downstream, at the interface of hardware instructions, which currently are always repeated for each new circuit. In this work, we modify this pipeline to compile only once to device-level instructions and instead perform inexpensive calls outside the \ac{qpu} kernel to update instruction lists or circuit parameters which can be quickly incorporated into the existing executables, essentially extending partial-compilation all the way into the kernel. This technique is outlined in Fig.~\ref{fig:intro}(b). By spinning up the \ac{qpu} exactly once per program we further reduce computational overhead for \ac{qci} problems. This, of course, assumes that the \ac{isa} of the control processor allows for partially compiled binaries. However, the novelty of our proposal lies in identifying a shared structure between execution iteration and utilizing this \ac{isa} to reduce compilation costs in \ac{qci} problems which involve repeated context switches between the classical and quantum computer. We also propose that this partial compilation at the device level can be extended to job execution queues for quantum computing in the cloud.

% \revision{The utility of this one-time \ac{dlpc} method depends on the need for hardware re-calibration and circuit re-optimization. For problems like traditional VQE, changes in parameters do not substantially affect the circuit optimizations. For problems where the structure does change iteratively, forced re-optimization every cycle has marginal benefits at the cost of another compilation loop and instead, this exit should be forced only periodically once changes have accumulated. Current hardware parameters do drift, requiring periodic re-calibration of the entire machine. This then necessitates circuit re-optimization (e.g. remapping) and kernel re-compilation since the hardware parameters have changed. In both these cases, we must recompile the kernel, limiting the effectiveness of the \ac{dlpc} technique. However, these events happen relatively infrequently, thereby still saving on compilation costs on most iterations. Also, given \ac{dlpc}'s application to calibration, and recent development in calibration routines that are dynamic and interleaved between computation (like in \cite{optimus}), our proposed technique shows an advantage over the naive approach even when the system needs to be re-calibrated and re-optimized}

% In this part you want to really hone in on what you do that's different and why its important - distill the message of the paper into 3-4 contributions
The major contributions of this work are as follows:
\begin{enumerate}
    \item We identify that quantum operations (gates or pulses) have invariant underlying structures at the machine level. 
    % For example, the pulse shape is fixed but its amplitude or frequency can be adjusted for different gate parameters, like rotation angle. 
    This enables parameterization of corresponding machine code which can dynamically adjust to changing circuit or pulse level parameters.
    \item We extend the idea of partial compilation to the device level, enabling one-time compilation of machine code run on the \ac{qpu}, reducing to nearly iteration-independent compilation scaling.
    \item We demonstrate this routine on a trapped-ion quantum computer by running simple \ac{vqe} programs and measuring the compilation and execution time speedup.
    \item DLPC reduces net runtime overhead for the class of \ac{qci} problems, or any iterative program, by demonstrating a potential speedup of up to 2.7x, 7.7x, and 2.6x for representative \ac{vqe}, optimal calibration, and \ac{rb} programs, respectively. \revision{For interleaved recalibration routines, we observe an even greater advantage of DLPC - we only compile the kernel for the circuits once but we also compile the kernel only once for the system probes and the calibration routines.}
    \item \revision{We generalize and expand DLPC to fixed gate-set hardware to accommodate cloud-based quantum platforms.}
\end{enumerate}

% The rest of this paper is structured as follows. \todo{describe sections}.

% \input{content/30_motivation}
\section{Background}
\label{sec:background}

\subsection{Compilation for QCI problems}
\label{subsec:compilation}
% { I think you can drop this paragraph} The hybrid nature of \ac{qci} programs make compilation for them an interesting problem. Every context switch between the \ac{cpu} and \ac{qpu} incurs a runtime overhead of re-compiling the quantum program  with a new set of parameters over each iteration. As the number of iterations increase, this compilation overhead can balloon out of proportion. However, the term compilation is used in the field as a blanket term for a multi-step process in the software pipeline. In the rest of this section, we break this compilation process into a few fundamental components, and highlight the one our proposal optimizes.

In a hardware-accelerator model of execution~\cite{svore2006layered, riesebos2019quantum, nguyen2020extending, chong2017programming}, a quantum program can be divided into classical components that run on the \ac{cpu}, and quantum components that run on the \ac{qpu}. The classical components are compiled by the high-level general-purpose language that they are represented in. The quantum components are compiled by the compiler built for the high-level domain-specific language used to represent a quantum circuit. We primarily focus on the latter, with quantum components being gate-level operations. Compiling quantum components in the \ac{nisq}~\cite{preskill2018quantum} era typically adheres to the following pipeline. The high-level gate instructions go through a hardware-agnostic optimizations,
% pass to simplify the circuit and remove any redundant operations. This is
followed by a compiler pass to transform the operations to a gate set specific to the target quantum hardware platform. Operations are then appropriately scheduled and mapped in relation to the hardware architecture. 
% In a trapped-ion system, the example hardware used in this work, this means determining which operations can be scheduled in parallel given all-to-all connectivity between qubits, and how the qubits are mapped to the trapped-ion chain considering ion shuttling requirements and individual ion error rates. 
This is followed by a conversion to pulses, which may be further optimized to reduce pulse length and account for known sources of error in the physical system. These pulses are then compiled to appropriate device-level instructions (essentially an assembly representation) which depend on the control electronics being used in the experimental setup. Finally, device-level instructions are translated into binaries which will be executed on the control electronics. 

% The entire process can be divided into 3 primary blocks - gate-level compilation, pulse-level compilation, and device-level compilation. 
% Gate-level compilation in turn can be broken down into a hardware-agnostic layer and a hardware-dependent layer. 
% In this work, we are focused on this final step - device-level compilation,and reducing redundant repetitions of this stage during \ac{qci} applications.

% Executing all the compilation routines described above over every iteration in a \ac{qci} routine can be computationally intensive, motivating the need to optimize this process.
% In this work we focus on device-level compilation - the overhead it adds, and propose a technique that reduces this overhead.

\subsection{Control System}

\label{subsec:control}
\Ac{artiq}~\cite{bourdeauducq_2016_51303, Kasprowicz:20} is a real-time control software solution that uses a hardware accelerator model of execution for quantum computing programs. The program can be offloaded to specialized hardware components allowing for greater efficiency as compared to running the program on a general purpose \ac{cpu} alone. In the case of \ac{artiq}, this specialized hardware component is a \ac{fpga} board optimized for real-time execution.
For the device-level compilation pipeline, we use an \ac{artiq} based control system~\cite{dax} to write and execute \ac{qci} programs. This control system is a modular software framework that builds upon \ac{artiq}. 

\begin{figure*}[h]
    \centering
    \includegraphics[width=\linewidth, scale=0.9]{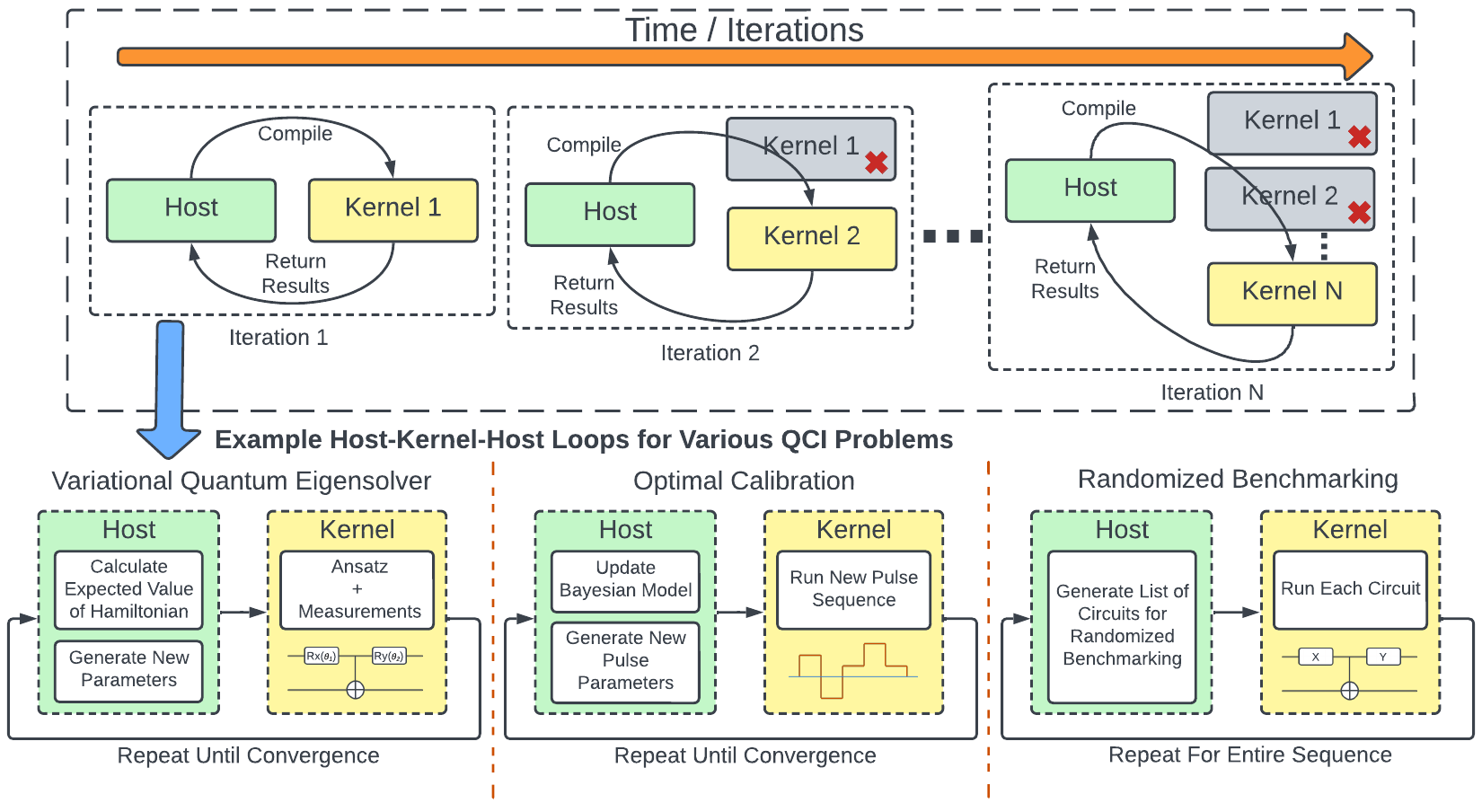}
    \vspace{-1.5\baselineskip}
    \caption{Typical compilation pipelines for \ac{qci} programs follow the control flow shown here. Each iteration of the program results in the compilation of a new kernel. 
    % The cost associated with this repeated compilation can add a significant execution time overhead. 
    Examples of host-kernel-host execution loops are shown. Over each iteration - in a \ac{vqe} program, the host generates parameters based on the expected value of the Hamiltonian; in an optimal calibration routine, the host generates new pulse parameters to converge to an optimal pulse sequence; in an \ac{rb} routine, the kernel executes a new circuit from a pre-generated list of benchmarking circuits. }
    % \vspace{-\baselineskip}
    \label{fig:qci_programs}
\end{figure*}

An \ac{artiq} program typically consists of 2 parts - \emph{host} and \emph{kernel}. The \emph{host} refers to the part of the program executed on a classical machine, which we refer to as the \ac{cpu}. The \emph{kernel} is performed on the real-time control \ac{fpga} board responsible for managing instructions on the quantum computer, which we refer to as the \ac{qpu}. For example, a classical optimizer is part of the host block while the real-time execution is defined in the kernel block of the program. The kernel is compiled to binary files to be run on the \ac{qpu} using \ac{artiq}'s compiler. The program on the \ac{cpu} communicates with the \ac{qpu} over an Ethernet connection.

Communication with the host from within the kernel can be done through \acp{rpc}. These \acp{rpc} can be synchronous, in which case the kernel waits for the call to return, or asynchronous where the call is executed on the host while the kernel continues with its execution. For example, storing the measurement results of a circuit is an asynchronous \ac{rpc}.

\subsection{QCI Problems}
\label{subsec:qci_problems}
Because we have hardware-level access to academic quantum systems, the description and demonstrations of \ac{dlpc} largely focuses on \ac{qci} problems in this paper. In particular, we focus on calibration, system characterization and \ac{vqe}. A sample schematic for these routines can be seen in Fig.~\ref{fig:qci_programs}.

\subsubsection{Optimized Calibration}
\label{subsec:calib}
Quantum computers require high fidelity (low error) state preparation, operations, and measurements. The parameters of a quantum system drift with time resulting in systematic errors that affect the fidelities. This motivates the need for periodic calibration of the system to appropriately adjust gate pulse parameters.
%Naive scan across parameters to re-calibrate the system can be expensive, and there have been recent proposals to make these calibration routines optimal~\cite{Majumder2020, PhysRevA.102.042611, Maksymov_2021, stace2022optimised}. 
These calibration routines involve quantum-classical interfacing requiring analysis and computation on the classical computer every iteration, searching for optimal pulse parameters such as in~\cite{stace2022optimised}.

\subsubsection{System Characterization}

\label{subsec:charac}
System characterization refers to routines that are run on a quantum computer to assess the noise in the system and are often used in conjunction with calibration. Widely used characterization routines include \ac{gst}~\cite{https://doi.org/10.48550/arxiv.1310.4492}, \ac{rb}~\cite{PhysRevA.77.012307, magesan2011scalable}, cycle benchmarking~\cite{Erhard2019}, and empirical direct characterization~\cite{dahlhauser2022benchmarking}. Most of these routines involve sequential execution of series of generated circuits, 
% and while they do not necessarily require this circuit execution to be interleaved with classical computation there is still
with a significant overhead associated with compiling large numbers of circuits each to be run only a few times on the quantum computer. We focus on \ac{rb} as an illustrative example of the present device-compilation overhead. \ac{rb} runs circuits with increasingly longer sequences of randomly generated gates \cite{PhysRevA.77.012307}.
% The general idea of \ac{rb} can be summarized as follows. For a given length, $l$, a series of circuits with $l$ gates are generated by randomly sampling from a set of gates operating on the number of qubits in the system. These circuits are then sequentially executed on the quantum computer, resulting in an effective unitary $U$. This is followed by a unitary $U^{\dagger}$ which brings the system back to its initial state \cite{PhysRevA.77.012307}. 
% This is repeated for increasing sequence lengths $l$. A comprehensive outline of this routine can be found at~\cite{PhysRevA.77.012307}. 
% Although this routine, along with other characterization algorithms, do not strictly fall into the class of \ac{qci} problems, they do involve repeated compilation of circuits. 
Similar to calibration, system operators characterize subsets, e.g. every pair of connected qubits, resulting in an additional factor of $O(N^2)$, $N$ being number of qubits, machine compilations~\cite{Helsen2019}.
% In the best case, it is preferable to sample \textit{more circuits} rather than \textit{more samples per circuit} meaning the actual time spent on the \ac{qpu} is relatively small.  

% Given that the circuits that are run as part of these algorithms are targeted primarily towards benchmarking, generated circuits are run as is, and the pulses that correspond to the operations in these circuits have been pre-optimized for each operation. This results in no significant gate and pulse-level compilation, with device-level compilation then being the bottleneck in their compilation pipeline. 
% In an example like \ac{rb}, where this routine can be repeated for sequence lengths on the order of $10^{2}$, the cost of compiling each circuit can be significant. 
% We discuss how our proposed technique reduces the device-level compilation overhead for characterization routines in Section~\ref{sec:fast}, and demonstrate the results of the speedup in Section~\ref{sec:results}.

\subsubsection{Variational Quantum Eigensolver}
\label{subsec:vqe}

\begin{figure*}[h]
    \centering
    \includegraphics[width=\linewidth]{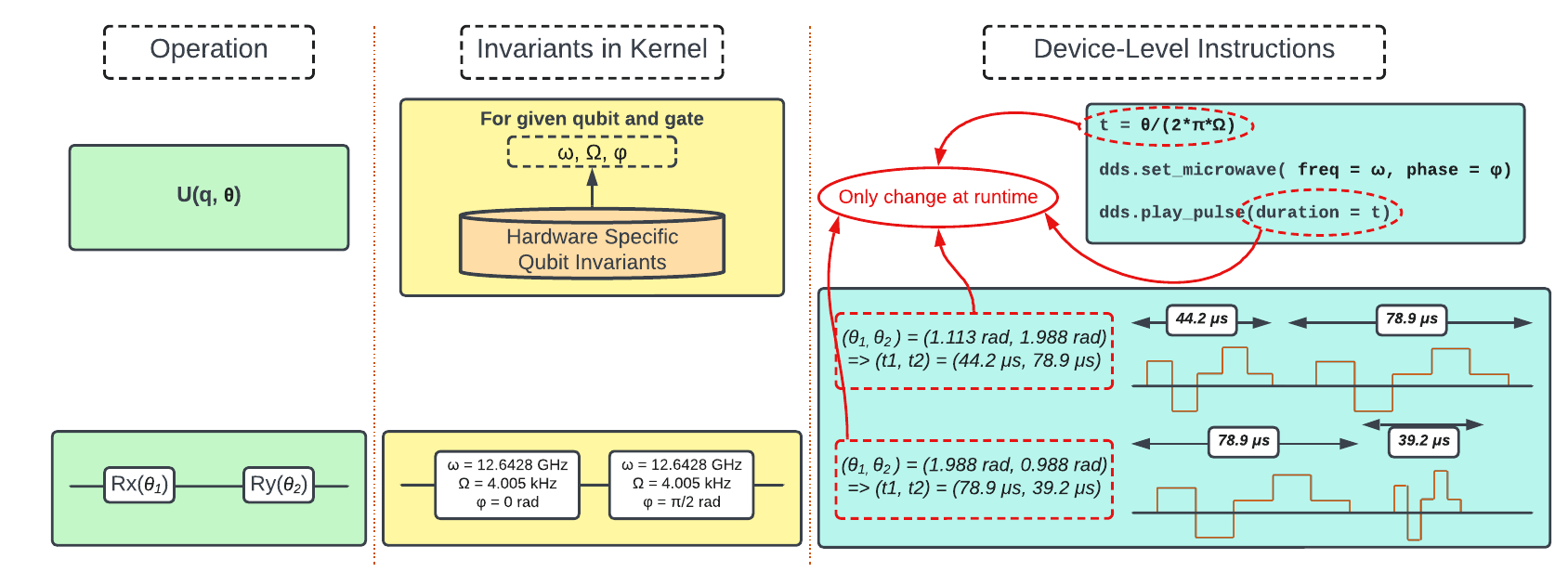}
    % \vspace{-1\baselineskip}
    \caption{Kernel execution of parameterized quantum operations. Consider a gate $U$ acting on qubit $q$ with a parameter $\theta$. First, the hardware-specific qubit invariants are retrieved from the persistent dataset. Here, invariant parameters are the qubit frequency $\omega$, Rabi frequency $\Omega$, and phase associated with the gates $\phi$. The desired pulse is configured by setting $\omega$ with a offset $\phi$ on the direct digital synthesizer (DDS), and $\Omega$ and the gate parameter $\theta$ are used to calculate the pulse duration, here from \cite{kim2020hardware}. The DDS plays the appropriate pulse sequence for the dynamically and analytically calculated duration. The kernel can be compiled just once with the invariant parameters, while the variables are supplied during runtime at each iteration.}
    
    % The single-qubit example here uses qubit parameters and pulse times from demonstrations on the quantum computer described in~\cite{kim2020hardware}. In this example, we consider a single-qubit parameterized circuit being repeated with different angles $\theta_{1}$ and $\theta_{2}$ at each iteration. It is observed that the only change in the executable at runtime is the pulse duration which is analytically computed quickly. This observation allows the kernel to be compiled just once with the invariant parameters, while the variables are supplied during runtime at each iteration, thereby saving on re-compilation costs.}
    % \vspace{-\baselineskip}
    \label{fig:kernel_flow}
\end{figure*}
For near-term quantum devices, a promising application is \ac{vqe}.
% This problem is considered to be exponentially hard to solve using a classical machine but can potentially be solved efficiently using a quantum computer. \ac{vqe} was developed as an alternative to \ac{qpe}, which is also capable of solving the same problem ~\cite{lloyd_qpe} but in a variational way rather than directly. Given the current error rates and number of qubits  available, \ac{qpe} is not a feasible algorithm in the \ac{nisq} era of quantum devices. For a given precision $\epsilon$, \ac{qpe} requires $O(1)$ repetitions of depth $O(1/\epsilon)$ circuits, while \ac{vqe} requires $O(1/\epsilon^{2})$ repetitions of depth $O(1)$ circuits ~\cite{Wang_2019}.
\ac{vqe} is a hybrid, iterative algorithm that uses the variational method~\cite{griffiths_schroeter_2018} and can be used to find the ground state energy of a molecule~\cite{Peruzzo2014}. 
% This is done by constructing a Hamiltonian whose minimum expectation value approximates the ground state energy of the molecule. 
% A Hamiltonian (essentially a large paramterized matrix) is a quantum mechanical operator which corresponds to the total energy of a system. A parameterized quantum circuit, or ansatz, with measurements corresponding to the constructed Hamiltonian is then iteratively executed with varying parameters until the expected value of the Hamiltonian converges to the optimal value.
On each iteration, the next set of parameters and set of measurements for the circuit are determined by a classical optimizer which runs on the \ac{cpu}, as opposed to the ansatz circuit, which is executed on the \ac{qpu}. %This routine of a classical optimizer working in tandem with the execution of quantum circuits makes \ac{vqe} a hybrid algorithm.%
At the gate-level, a \ac{vqe} program remains invariant with respect to the type of gates being used across iterations. However, the changing parameters of these gates over each iteration have cascading effects on the resultant pulses and device-level code. In the naive pipeline, this results in these components being re-compiled every iteration. 
% Past work, like \cite{gokhale_variational}, propose compiler optimizations and partial compilation techniques to reduce the pulse-level re-compilation overhead. 
% In Section~\ref{sec:results}, we show that for real small-scale systems the DLC time can be significant, and demonstrate how our proposed model reduces this overhead to being nearly constant.
\section{Control Flow for QCI Routines}
\label{sec:current}

Currently, the control software design for a \ac{qci} program 
%is as follows. The control flow%
begins in the host. In an optimal calibration routine, this is where the initial pulse parameters are declared, while in a \ac{vqe} program the initial guess for the ansatz parameters are declared here. This is followed by an invocation of a kernel function that runs the pulse sequence or the ansatz circuit respectively on the quantum system with the given parameters. The control is then transferred back to the host. 
% \revision{Here},
% For the optimal calibration routine described in~\cite{stace2022optimised} this is where 
Based on measurement results 
% post the calibration routine and a 
the classical optimizer generates a new set of parameters for the next iteration. 
% For a \ac{vqe} program,based on the Pauli operators measured by the ansatz,
% the expectation value of the Hamiltonian is calculated. If convergence has not been reached (defined by the user), the process is repeated with a new set of parameters generated by the classical optimizer. This control flow can be seen in Fig.~\ref{fig:qci_programs}. This may require large numbers of iterations.

% Typically users use off-the-shelf optimizers from libraries like Scipy~\cite{2020SciPy-NMeth} which
% % These optimization functions have arguments that include the optimizer method to use and the objective function. This objective function 
% take in the current parameters as arguments and calculates the cost based on these parameters. Internally the optimizer repeatedly calls the objective functions, generating new parameters each time, until the convergence is reached. When using these optimizers, the user defines the optimizer method, the initial parameter guess, and the objective function. For example, in a \ac{vqe} program, this objective function first runs the ansatz circuit based on the current parameters passed in as arguments and uses the measured results to calculate and return the expected value of the Hamiltonian being minimized. This is repeated until the expectation value of the Hamiltonian converges. Our work does not modify any of these steps. A similar structure applies for other iterative applications.

In the existing control flow, each iteration of the \ac{qci} routine calls a kernel function, as seen in Fig.~\ref{fig:qci_programs}. Each call involves device-level compilation passes which have a runtime cost associated with them. This cost consists of kernel compilation, kernel uploading, and kernel scheduling. Kernel compilation involves the time it takes to optimize and compile the kernel code to be executed on the control hardware. This depends on the complexity of the kernel. Kernel uploading is the time it takes to upload the compiled kernel onto the control hardware. This is determined by the size of the compiled kernel and the communication time between the host and control hardware. Finally, kernel scheduling consists of the time it takes to schedule the compiled kernel on the control hardware execution queue. This is affected by whether there are any other processes that need to be terminated before the current one is allowed to execute.

All of these components repeated in each iteration affect the runtime performance of a \ac{qci} program. We identify an inherent underlying shared structure in the kernel code of \ac{qci} programs, as in Fig.~\ref{fig:kernel_flow}. For example - \ac{vqe} programs always execute the same circuits, only varying the gate parameters at each iteration; an optimal calibration routine fixes a general structure for the pulse sequence while varying a subset of the pulse parameters, like the frequency and the number of segments in a piece-wise segmented sequence remain fixed in~\cite{stace2022optimised}, while the amplitude and total pulse duration change every iteration; a \ac{rb} routine runs a sequence of circuits for varying number of gates per circuit but all of these gates are randomly sampled from a set pool of operators which can be precompiled. This underlying structure results in large parts of the kernel being \textit{invariant} at runtime. This invariance lends the device-level code to be partially compiled, with dynamic parameters
% , like output channel or pulse amplitude, 
passed in at runtime. The following section describes our \ac{dlpc} technique that saves on the repeated kernel compilation, uploading, and scheduling time over each iteration of the program.
% In this work, we've identified low-level redundancy that ignores fundamental properties of operations executed on quantum hardware - namely that when changing only the angle parameters of gates in the circuit, or changing the circuit when it is based off a set pool of operators with no optimization, or when changing only certain parameters of a pulse sequence, the executable can be easily modified. The following section describes our partial device-level compilation technique that  saves on these kernel compilation, uploading and scheduling time that are repeated over each iteration of the program.

\section{DLPC}
\label{sec:fast}

\begin{figure}
    \centering
    \includegraphics[width=\linewidth]{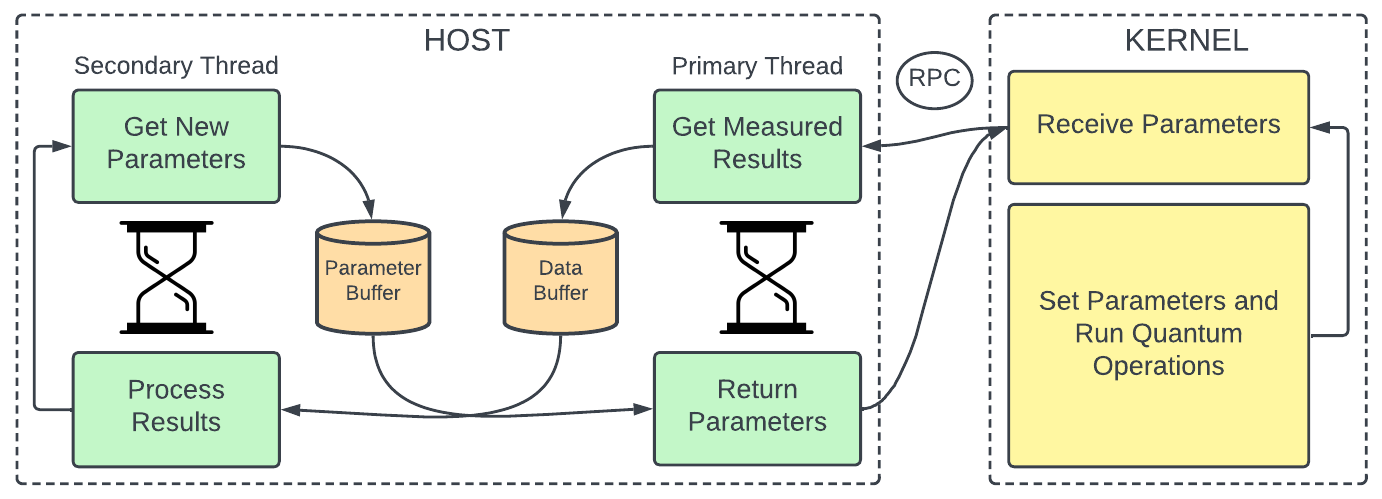}
    % \vspace{-1\baselineskip}
    \caption{Our proposed partial-compilation routine for \ac{qci} programs. \revision{The kernel is compiled only once, and at runtime receives new parameters through \acp{rpc} at each iteration. The host program uses a multi-threaded produces-consumer pattern to wait on the latest measurement results and appropriately generates new parameters or sends a sentinel value conveying convergence.}}
    \vspace{-1\baselineskip}
    \label{fig:new_impl}
\end{figure}

In principle, the simplest way to optimize device-level compilation is to avoid re-compiling blocks of kernel code every iteration. We propose a routine that has a kernel block compiled only once and gets new parameters from the classical computer at each iteration. 
% This, in effect, results in partial compilation of the kernel.

There are some constraints that need to be considered while designing this routine. Consider the example of a \ac{vqe} program - once the program enters the kernel to execute the first iteration of the ansatz, the control flow has to stay in the kernel until the last iteration of the ansatz is run. Failure to do so results in re-compilation of the kernel. The classical optimizer has to run on the host, as the hardware on the \ac{qpu} lacks the ability to perform computationally intensive tasks. This requires the control flow in the kernel to wait for updated ansatz parameters between iterations. 
% Furthermore, the routine must be compatible with off-the-shelf optimizers.
% As outlined in Section~\ref{sec:current}, these optimizers allow users to define an objective function. Similar to the control flow waiting in the kernel for new parameters between iterations, the control flow in this objective function also needs to wait for the measurement results from the circuit. Similar constraints hold for the optimal calibration routines.

Given these constraints, we propose the following \ac{dlpc} routine. On the \ac{cpu}, we divide the execution of the program into two threads - the main thread that runs the primary control flow of the host, and a secondary thread that runs the optimizer function. These threads wait on 2 buffers respectively - a parameter buffer and a results buffer. The parameter buffer stores the new parameters for the next iteration of the \ac{qci} program, and the results buffer stores the latest measurement results from the operations executed on the quantum system. In this routine, these two threads will work similar to a multi-threaded producer-consumer pattern. The \ac{qpu} consists of just one thread for kernel execution. The \ac{qci} routine starts with the secondary thread waiting on the data buffer and the main thread calling the kernel function for the first time with initial parameters. This is the only time the kernel is compiled and uploaded on the \ac{qpu}. 

In the case of the VQE example, the kernel function is terminated only when the optimal expectation value is reached. In the first iteration, the kernel executes the quantum operations with the initial parameters and then makes a synchronous \ac{rpc} to retrieve the next set of parameters. Because this is a synchronous \ac{rpc}, it waits for the call to return before resuming execution. This \ac{rpc} invokes a function on the main thread which first adds the latest measurement results into the data buffer and then waits on the parameter buffer. This resumes execution on the secondary thread running the optimizer's objective function. This secondary thread performs the intensive classical computation and analyses based on the measurement results, generates a new set of parameters, adds them to the parameter buffer, and then goes back to waiting on the data buffer. This resumes execution on the main thread that was waiting on the parameter buffer. The function on the main thread returns these new parameters back to the kernel, which resumes its execution by running the quantum operations again with the new parameters, appropriately adjusting pulse amplitudes, frequencies, etc. The routine continues until the optimizer converges, at which point it adds a sentinel value to the parameter buffer. This results in the main thread returning a Boolean flag back to the kernel that terminates its execution. This routine is illustrated in Fig.~\ref{fig:new_impl}. This partial-compilation routine can be adapted to any \ac{qci} routine, like characterization or optimal calibration, as they follow the general structure of interleaving execution on the \ac{qpu} with some amount of computation on the \ac{cpu} to inform the next iteration.

Here the kernel is compiled to the \ac{qpu} only once but is not yet executable as the parameters for the quantum operations are variable. Over each iteration of the \ac{qci} program, this partially compiled block gets the latest parameters which are appropriately added to the compiled binaries and then executed on the \ac{qpu}. The only overhead for each iteration of the program is the communication round-trip time taken by the synchronous \ac{rpc}. However, this communication round-trip time is also incurred by the existing device-compilation pipeline over every iteration but is generally negligible, in addition to the time taken to compile, upload and schedule the kernel block each time. The multi-threaded producer-consumer design pattern using two shared data structures allows users to continue using off-the-shelf optimizer functions while avoiding race conditions.

% A similar technique can be adapted to benchmarking algorithms that cannot be strictly classified as \ac{qci} problems. These programs do not necessarily require the execution of quantum operations to be interleaved with classical computation, though they could if there is any decision-making about subsequent experiments, for example, to increase precision. Regardless, they do follow an iterative structure of running a series of circuits sequentially. Compiling each of these circuits can add a significant performance overhead as these benchmarking routines could consist of a large number of randomized circuits~\cite{PhysRevA.77.012307}. A vital distinction of these routines is, however, that their circuits are built from a set pool of operators and do not go through any optimization procedures as they are meant to assess the current state of the quantum system. 

The kernel block for other routines, like characterization, can be written to only include functions that correspond to operators in the pool. The kernel can be parameterized to receive a simple representation of the circuit, and at runtime the circuit can be executed by appropriately calling the function that corresponds to the gates in the circuit representation. As each circuit is composed of the same set of gates, this results in the kernel being compiled only once. Depending on the memory available on the \ac{qpu}, a block of circuit representations that parameterize the kernel can be uploaded through a \ac{rpc}, much like the parameters for \ac{qci} routines. This partial-compilation technique saves on the overhead added by compiling each circuit of the sequence by leveraging their structure. Given that the number of circuits executed as part of these benchmarking routines can be very large, our one-time compilation technique offers a significant performance benefit. This can be extended to hardware with long periods of uptime, e.g. commercial hardware accessed via the cloud. \revision{This is explored in Sec~\ref{subsec:beyond_qci}}.

% The novelty of the proposed \ac{dlpc} routine lies in identifying the shared structure between iterative runs of \ac{qci} problems and building a device-level partial compilation pipeline that uses an always-on kernel and producer-consumer multi-threaded design pattern on the host. This pipeline utilizes the shared structure in these problems with a control processor whose \ac{isa} supports partially-compiled executable binaries.
% The following sections discuss the speedup of our partial device-level compilation routine in relation to the existing compilation technique.
The novelty of the proposed \ac{dlpc} routine lies in identifying the shared structure between iterative runs of \ac{qci} problems and building a device-level partial compilation pipeline that uses an always-on kernel and producer-consumer multi-threaded design pattern on the host. Our work does not present changes to the compiler or the \ac{isa}, which has been done previously \cite{Karalekas_2020}. The pipeline utilizes the shared structure in \ac{qci} problems in tandem with an underlying architecture that supports partially compiled executable binaries to compile to the kernel only once and stream the execution of subsequent programs without re-compilation.

\section{Results}
\label{sec:results}

To demonstrate the \ac{dlpc} technique on an experimental system we use a trapped-ion quantum system at Duke University described in~\cite{kim2020hardware}. We used this academic system as we needed low-level control access to the system for our demonstration.
% It is a cryogenic trapped-ion system that is designed to operate on a chain of 32 ions using a multi-channel \ac{aom}. 
%At the time of this demonstration, the system was only capable of running operations on a single-qubit. 
We demonstrate the pipeline on  single-qubit \ac{vqe} problems, but the \ac{dlpc} routine can be used on any number of qubits and operations. We use experimental results to demonstrate our proposal's immediate viability. 
% and use in both academic and potentially industrial quantum computers.

In order to show the performance of the partial-compilation technique on \ac{qci} problems involving multiple qubits, we use a functional simulator~\cite{dax:sim} that gives us accurate compile times for the kernel. The runtime of these routines is then estimated using true operation times on the trapped-ion quantum computer described above in addition to the compile time from the simulator. These operation times are $5 \mu s$ for single-qubit operations and $150 \mu s$ for two-qubit operations. In all these results \emph{runtime} refers to the total runtime of the program from when it is submitted to the end of execution. \emph{Compile time} refers to time taken to compile the device-level ARTIQ program, upload the binary to the board, and schedule it for execution. In our profiling, the latter two are insignificant relative to the compilation time. Lastly, \emph{kernel time} refers to time spent executing the binary on the kernel, this includes any computation done on the kernel and runtime on the physical quantum system.

\subsection{Experimental Results}
\label{subsec:experiment}

\begin{figure*}[h]
    \centering
    \includegraphics[width=\linewidth]{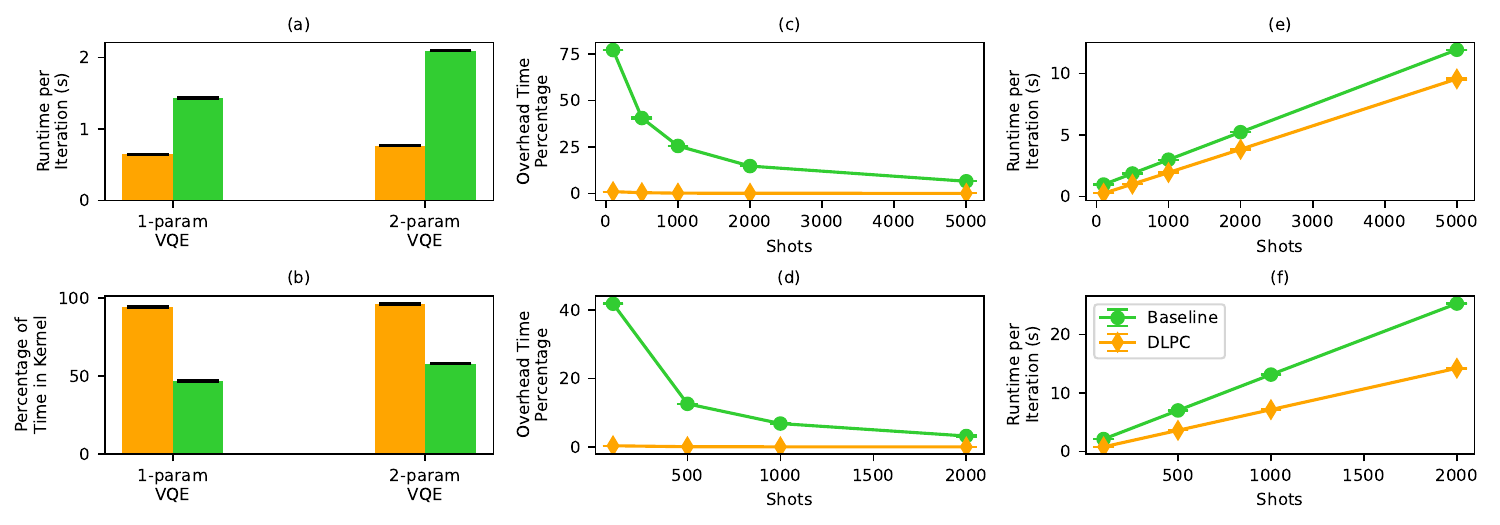}
    \caption{\revision{Demonstration of \ac{dlpc} technique on a trapped-ion quantum computer. (a) Runtime comparison between the baseline and \ac{dlpc} approaches. On average, the baseline approach takes 2.2x more time than the DLPC technique for the 1-param \ac{vqe} demonstration, and 2.7x more for the 2-param \ac{vqe} demonstration. (b) Percentage of time spent executing the circuit on the quantum computer. On average, the \ac{dlpc} technique results in 94.4\% of time spent in the kernel, relative the baseline's 46.8\% for the 1-param \ac{vqe}. For the 2-param \ac{vqe}, on average 96.1\% of time is spent in the kernel when using \ac{dlpc} relative to the baseline's 57.9\%. (c)-(d) Compilation overhead as a percentage of total runtime for increasing number of shots for the 1-param and 2-param \ac{vqe} experiments respectively. (e)-(f) The runtime per iteration as a function of the number of shots for the 1-param and 2-param \ac{vqe} experiments respectively.} }
    % The relatively larger difference in the slopes of the baseline and \ac{dlpc} approaches for the 2-param \ac{vqe} experiment can be attributed to larger compilation overhead.}}
    \label{fig:vqe_plot}
\end{figure*}

For the experimental demonstration of the \ac{dlpc} routine, we execute 2 single-qubit \ac{vqe} programs.
% as simple (not corresponding to any real molecule) but useful demonstrations of the quantum-classical execution model. 
%The first program uses an ansatz with one parameterized gate and a $\sigma_{z}$ measurement. The second one has an ansatz with two parameterized gates followed by $\sigma_{x}, \sigma_{y},$ and $\sigma_{z}$ measurements on the qubit. 
For consistency, both the ansatz circuits are sampled over 300 shots for each iteration of the \ac{vqe} program. In the single-parameter circuit, this results in 300 runs of the circuit followed by a $\sigma_{z}$ measurement, while in the two-parameter circuit, it involves 100 runs each of the circuits followed by $\sigma_{x}, \sigma_{y},$ and $\sigma_{z}$ measurements respectively. In both cases, the data has been collected over 10 runs of the \ac{vqe} program. The results from this demonstration are presented in Fig.~\ref{fig:vqe_plot}.

When sampling over 300 shots, the naive compilation approach takes, on average, $\sim$2.2x  more time for the 1-parameter \ac{vqe} and $\sim$2.7x for the 2-parameter \ac{vqe}. The latter shows a larger speedup as it involves a bigger kernel as kernel compilation is related to the number of gates.
% with more instructions resulting in a longer kernel compilation time; that is the kernel compilation overhead is directly related to the number of uniquely defined gates, and in the baseline this extends to gates which only differ in their parameters, such as angles. The cause of this speedup is apparent 
\revision{In Fig~\ref{fig:vqe_plot}(b)} we show the percentage of time each algorithm spends in the kernel executing the circuit. For the naive case, this percentage is $\sim$46.8\% for the 1-parameter \ac{vqe} experiment and $\sim$57.9\% for the 2-parameter \ac{vqe} experiment, while when using our technique it is $\sim$94.4\% and $\sim$96.1\% respectively. When using the naive approach, a large portion of the execution time is spent in re-compiling the kernel at the device level. However, in applications like \ac{vqe}, as the size of the problem and the required number of shots increase, the execution time is dominated by the circuit time~\cite{Gonthier_2022}. \revision{This results in a diminishing advantage of our compilation technique, which has been demonstrated in the Fig~\ref{fig:vqe_plot}(c)-(d)}. In the worst case, there is always some constant advantage with our technique, but the extent of this advantage depends on the \ac{vqe} techniques~\cite{Grimsley2019} used and the ongoing research in improving gate times~\cite{Schäfer2018} and reducing \ac{vqe} shot budgets~\cite{gu_shot_budget}. Our proposal's real strength is applications that require many different circuits but which get run relatively few times.

\subsection{Simulation Results}
\label{sunsec:simulation}

We demonstrate the performance of our \ac{dlpc} technique for optimal calibration routines and system characterization schemes by running these routines in simulation to get true compile times and estimated runtimes of the experiments.

\begin{figure*}[h]
    \centering
    \includegraphics[width=\linewidth]{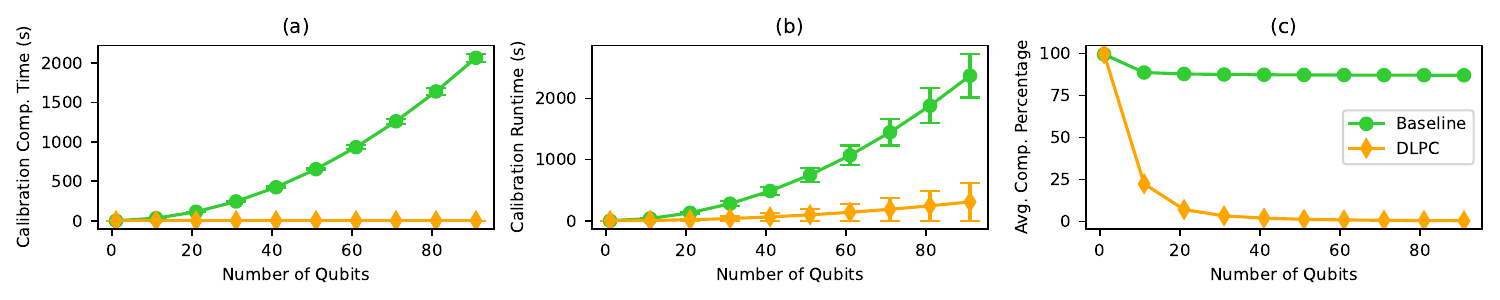}
    \caption{\revision{Demonstration of the \ac{dlpc} technique relative to the baseline approach for the optimal calibration routine proposed in~\cite{stace2022optimised}. (a) \ac{dlpc} compiles the kernel only once, resulting in a constant compilation cost of 1.2 seconds on average. The baseline approach compiles the kernel for every single iteration of the routine which scales quadratically. (b) The runtime for \ac{dlpc} and baseline approaches grows quadratically as the system size increases, however, the baseline approach grows much faster due to the dominant compilation overhead. (c) The compilation time as a percentage of the total runtime. For the \ac{dlpc} approach, this percentage diminishes as the system size grows, while in the baseline approach, this is over 80\%. }}
    \label{fig:calib_plot}
\end{figure*}

Fig.~\ref{fig:calib_plot} presents the performance advantage of the \ac{dlpc} technique for the optimal calibration procedure proposed in ~\cite{stace2022optimised}. Their proposal has a calibration routine for single-qubit pulses using 2 parameters and for two-qubit routines using 5 parameters. For $N$ qubits, the total number of parameters can then be represented as $2N + 5\binom{N}{2}$. 
% This is because a $N$ qubit system lends itself to single-qubit operations on any of the $N$ available qubits, and two-qubit operations on a combination of any 2 of the $N$ available qubits. 
The number of iterations required for convergence is proportional to the number of parameters in the system, resulting in a superlinear scaling of iterations with qubit size. The parameterized kernel for this routine takes in the qubits being optimized, the amplitude modulations, and the pulse duration as parameters for the piece-wise constant pulse. Our technique compiles this kernel to the device only once, while in the naive approach, this compilation cost has at least quadratic scaling. Their calibration routine claims to converge to the optimal routines using 50-100 shots per experiment. Given single and two-qubit pulses on the order of the single and two-qubit gate times mentioned above, the total runtime of the experiment is then dominated by the compilation time. This results in the total runtime scaling similar to the compilation time of the routine.

\begin{figure*}
    \centering
    % \vspace{-0.7\baselineskip}
    \includegraphics[width=\linewidth]{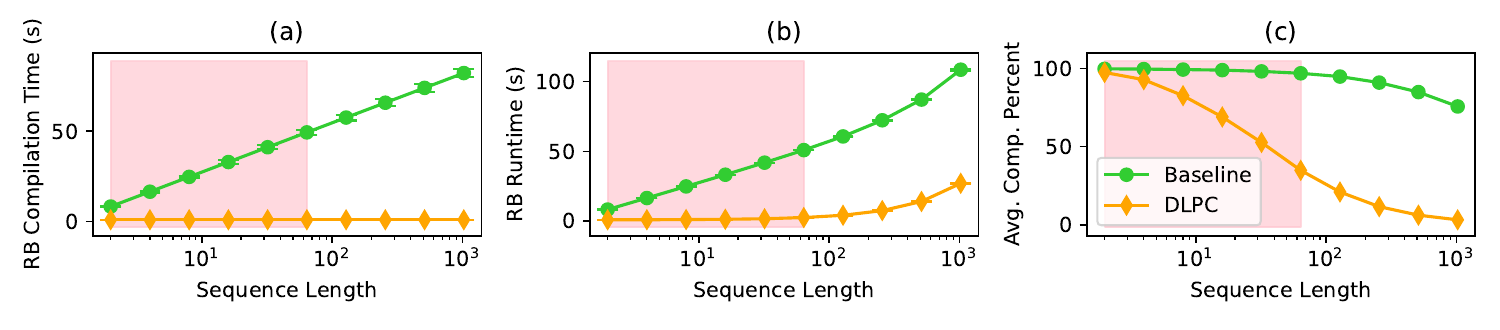}
    % \vspace{-1.5\baselineskip}
    \caption{\revision{\ac{dlpc} relative to the baseline approach for \ac{rb}. The shaded region indicates the typical sequence lengths for current machines which do not exceed $\sim10^{2}$ (a) The \ac{dlpc} routine results in an average constant compilation cost of 0.87 sec. For the baseline approach, compilation cost scales with a growing sequence length. (b) The runtime for \ac{dlpc} and baseline approaches grows as the number of \ac{rb} sequences increase, however, the baseline approach grows much faster due to the dominant compilation overhead. (c) The compilation time as a percentage of the total runtime. For \ac{dlpc}, this percentage diminishes as the sequence length increases, while for the baseline it is over 75\%.}}
    % \vspace{-0.7\baselineskip}
    \label{fig:rb_plot}
\end{figure*}

Fig.~\ref{fig:rb_plot} similarly demonstrates our technique for the \ac{rb} system characterization routine. The parameterized kernel here consists of the native gate set that makes up every circuit in the characterization routine and takes in the sequence of gates to be run every iteration as parameters. 
% The size of the kernel here only depends on the native gate set of the physical system.
% The plots in Fig~\ref{fig:rb_plot} use a logarithmic scale  for the x-axis as \ac{rb} sequence lengths increase as powers of 2. 
% In these plots, for each sequence length $2^{n}$, \ac{rb} routines have been executed for sequence lengths $2^{1}, 2^{2},...,2^{n-1}$. 
The demonstration executed 10 circuits for each sequence length. The naive compilation approach re-compiles this for every circuit, resulting in a compilation that grows with the number of circuits and number of gates in each circuit being executed on the quantum computer. Here each circuit is repeated for 100 shots, resulting in a runtime that is dominated by the compilation time for typical sequence lengths (which are at most on the order of ~$10^{2}$).
Consequently, the total runtime for the \ac{rb} routine scales much faster when using the naive approach relative to our proposed technique.
DLPC's gain would be further amplified on \ac{rb} protocols~\cite{Harper_2019, Granade_2015} that randomize the circuit run for each shot.
% , which is also reflected in the percentage of total runtime spent compiling. 
The demonstration here only explores scaling \ac{rb} with its sequence length, however, its scaling with system size shows a similar trend to that of the calibration routine. This is because, similar to calibration, characterization routines also characterize subsets of qubits, resulting in an additional factor of $O(N^2)$ machine compilations.
% It is interesting to note that in Fig.~\ref{fig:rb_plot}(c), the compilation time percentage data demonstrates an inflection point where the compilation cost begins grows at a rate slower than the circuit execution cost.
\section{\revision{Generalizations}}
\label{sec:gen}

\subsection{\revision{Beyond Trapped Ions}}
\label{subsec:beyond_ti}

\begin{figure}[h]
    \centering
    \includegraphics[width=\linewidth]{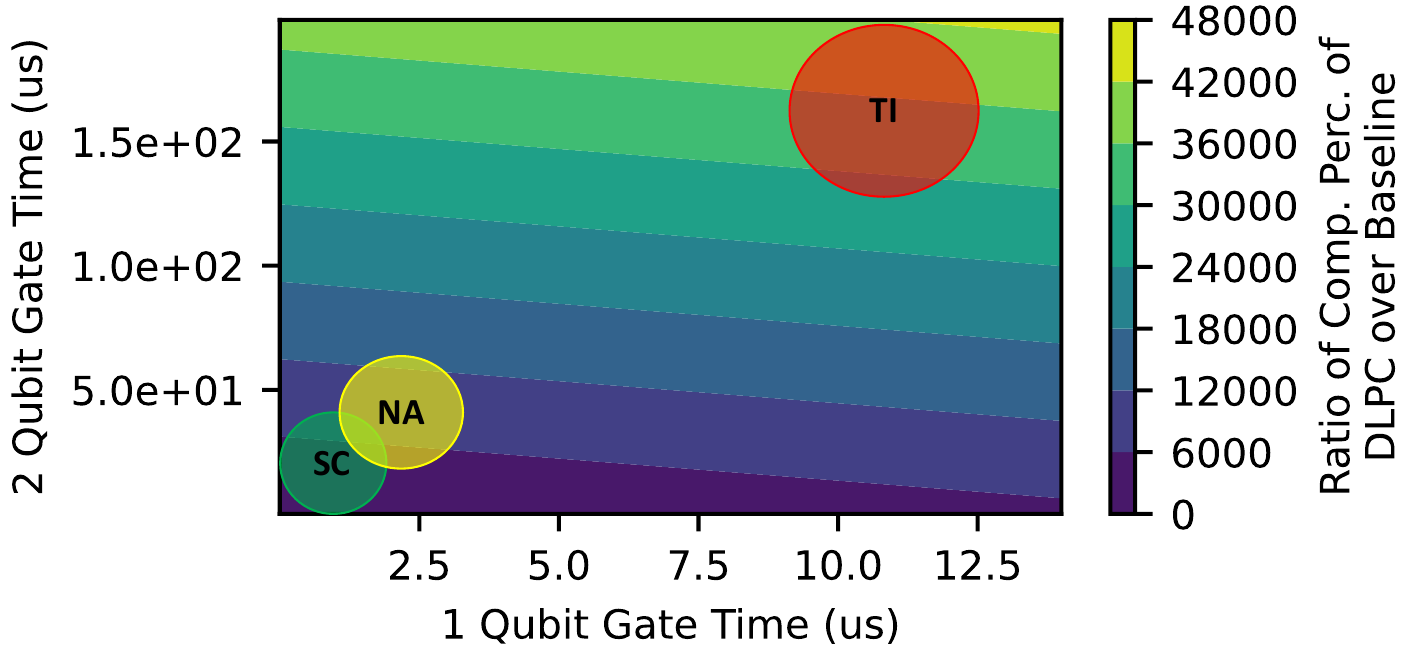}
    % \vspace{-1.5\baselineskip}
    \caption{\revision{Evaluation of \ac{dlpc} across machines. The contour plots the ratio of the compilation percentage of \ac{dlpc} to the baseline approach as a function of gate times. Compilation percentage is given by compilation time over total execution time. A small number on the contour indicates that the baseline approach spent a large portion of its runtime compiling relative to \ac{dlpc}. The figure indicates typical runtimes of superconducting (SC)~\cite{Kim2023}, neutral atom (NA)~\cite{Graham2022} and trapped-ion (TI)~\cite{Bruzewicz_2019} systems, with TI systems demonstrating the least advantage and SC the most.}}
    % \vspace{-1\baselineskip}
    \label{fig:vqe_machine}
\end{figure}

\revision{The results discussed in the previous section were demonstrated using trapped-ion machines with fixed gate times. Operation times play a crucial role in the relative advantage provided by the \ac{dlpc} technique; the speedup from DLPC diminishes if the runtime of the circuit on the quantum system dominates the execution time of the program. This circuit runtime is a factor of the gate times on the physical hardware. In Fig~\ref{fig:vqe_machine} we demonstrate \ac{dlpc} speedup as a function of gate times per iteration for a 4-qubit \ac{vqe} example run over 40000 shots and 100 iterations. Here we represent the ratio of the fraction of total program runtime spent compiling the kernel for the \ac{dlpc} technique relative to the baseline approach. A smaller number indicates that the baseline compilation method spends a longer time compiling relative to \ac{dlpc}.}
% For faster gates, like those seen in superconducting machines, \ac{dlpc} provides a bigger advantage relative to the baseline approach, as compared to systems with slower gate times.}

\subsection{\revision{Beyond \ac{qci} Routines}}
\label{subsec:beyond_qci}

\revision{While primarily motivated by \ac{qci} problems, \ac{dlpc} can also be extended to non-\ac{qci} programs as demonstrated in the \ac{rb} target application. A potential non-\ac{qci} application of \ac{dlpc} is quantum computers in the cloud. Most commercial quantum computers are available to users through the cloud~\cite{ravi2022quantum}. These machines follow a workload management system where users submit their quantum jobs to a queue.
%and these are executed on the system on a first-come-first-serve basis. 
Each circuit is first compiled and optimized at the operation level, before being compiled and uploaded to the \ac{qpu} for execution. However, given that each of the submitted circuits to the queue is targeting the same machine, they have an underlying shared structure of using the same native gate set. This shared structure can be used to extend \ac{dlpc} to the quantum computing workflow in the cloud.}

\begin{figure}[h]
    \centering
    % \vspace{-0.3\baselineskip}
    \includegraphics[width=\linewidth]{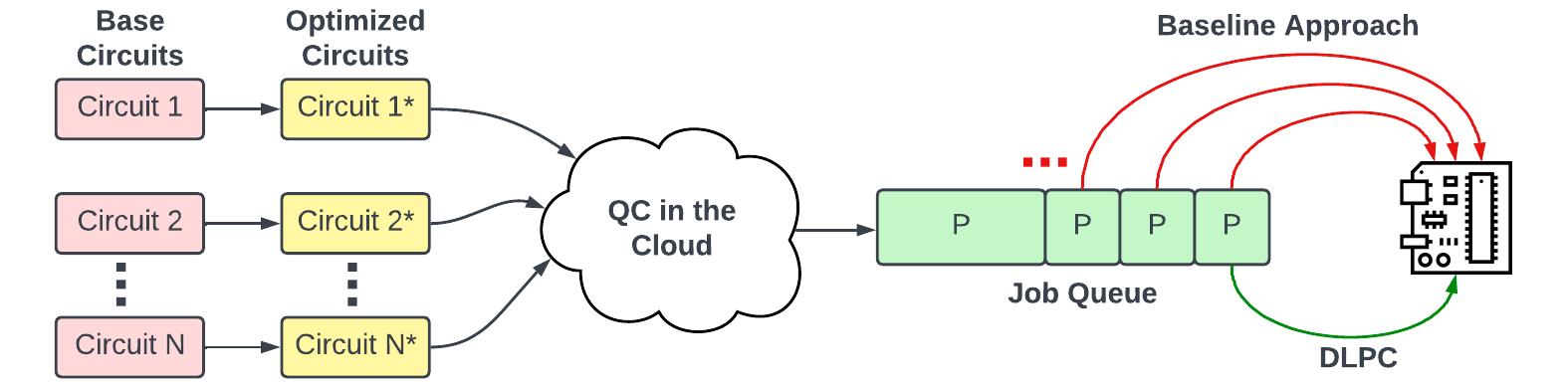}
    % \vspace{-\baselineskip}
    \caption{\revision{Workflow for quantum computers hosted on the cloud start with users submitting their circuits. These circuits are optimized on the client side before being added to a job queue. The baseline approach compiles the kernel for each of these circuits prior to execution. \ac{dlpc} re-compiles the kernel only when the jobs in the queue go to zero, or the system needs to recalibrate.}}
    % \vspace{-\baselineskip}
    \label{fig:qc_cloud_infra}
\end{figure}

\revision{Each circuit submitted to the quantum computer still needs to be compiled and optimized at the circuit level. Once it has been added to the execution queue and reduced to the native gate set, \ac{dlpc} can be used to save the repeated device-level compilation costs. A simple way to ensure that there is no kernel re-compilation and that all the native gates have been pre-compiled on the system is to run a dummy circuit consisting of all the native gates at the beginning of the queue. Fig~\ref{fig:qc_cloud_infra} summarizes the design of the \ac{dlpc} technique applied to quantum computers in the cloud.}

\begin{figure*}[h]
    \centering
    \includegraphics[width=\linewidth]{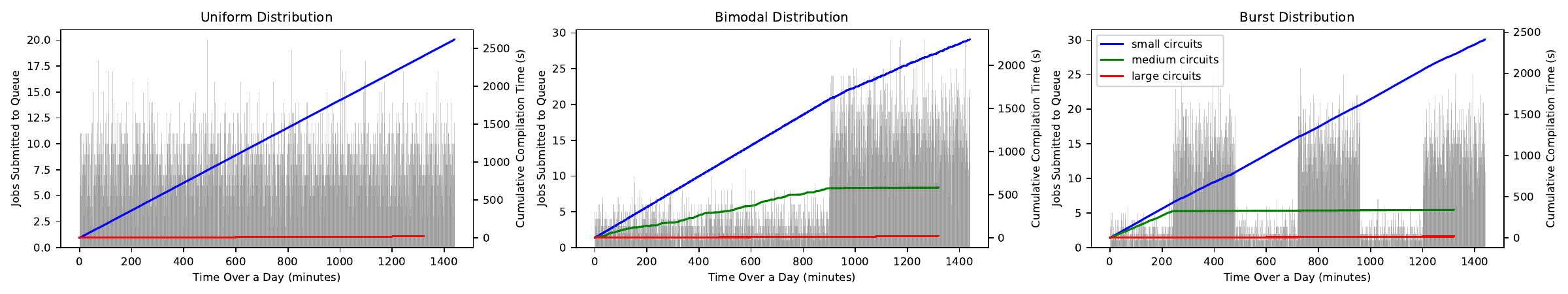}
    % \vspace{-1\baselineskip}
    \caption{\revision{Cumulative kernel compilation time for \ac{dlpc} over a day across 3 distributions - uniform, bimodal and burst; and 3 circuit sizes - small ($\sim$30 gates), medium ($\sim$100 gates) and large ($\sim$200 gates). %Small circuit sizes have the largest cumulative time, followed by medium, and finally large circuit sizes result in kernel re-compilation only when the system needs to be re-calibrated.% 
    In comparison, the baseline approach re-compiles for each circuit resulting in cumulative compilation cost of $\sim$2.7 hours over a day.}}
    % \vspace{-\baselineskip}
    \label{fig:qc_cloud}
\end{figure*}

\revision{In Fig~\ref{fig:qc_cloud} we emulate cloud workflow over a day to approximate the cumulative time spent compiling circuits using \ac{dlpc}. The kernel only recompiles if the number of jobs in the queue go to zero, or the system needs to undergo a re-calibration procedure (considered to be every 2 hours). We demonstrate results for 12000 circuits a day with varying workflow distributions through the course of a day and varying circuit sizes. Small circuits result in the largest cumulative kernel compilation time ($\sim$40.7 minutes across distributions) as the jobs in the queue are executed quickly, resulting in the queue going to zero routinely. 
% This is followed by medium circuits, and finally by large circuits. 
For large circuits, long execution times result in the kernel being recompiled only for periodic re-calibration leading to a minimal compilation cost. In contrast, the baseline approach recompiles the kernel for each circuit resulting in a cumulative average compilation time of $\sim$2.7 hours over a day.}

\section{\revision{Re-optimization and Re-calibration}}
\revision{In prior sections, we have assumed that both \circled{1} changes in circuit parameters does not require a re-optimization of the resulting circuit and \circled{2} device drift is insignificant. In both cases, this advantages DLPC because the compiled information in the kernel is assumed perfect. In practice, neither is a reasonable assumption; changes in gate parameters can possibly be optimized as they change
% , for example exposing possible gate cancellations, 
and device parameters can drift resulting in stale device level instructions. 
% (for example a change in the frequency of a qubit requires new gate definitions).
}

\subsection{\revision{Circuit Re-optimization}}

\revision{Changes in gate parameters, such as rotation angles, do not often result in significant structural changes in the circuit, even with a complete circuit-level re-transpilation. If error rate drifts could be accurately predicted, high level circuit optimization could dynamically remap programs onto less error prone qubits; no such accurate prediction exists today.}

In typical \ac{vqe} programs, re-optimizing and re-compiling the circuit between iterations results in some gate count and circuit duration changes. These changes, however, are due to non-determinism in mapping and routing algorithms and not because of significant circuit re-optimization. If the transpilation is seeded, these variations vanish. Frequent exits out of the host-kernel loop are largely unnecessary, and instead, it is more favorable to highly optimize the original parameterized circuit and then require only a single kernel compilation. \revision{In some implementations of variational algorithms, e.g. ADAPT-VQE \cite{Grimsley2019}, the circuit structure itself is a variational parameter and different circuit substructures are added at each iteration. Circuit re-optimization could result in significantly reduced depth or gate count, both the primary sources of error. However, re-compiling every iteration adds minimal gain, and it is more practical to periodically force re-optimization of the circuit which will also require a recompilation of the kernel. If there are significant circuit changes warranting re-optimizations every iteration, \ac{dlpc} would still accommodate this due to the underlying shared structure of the native gate set (see Sec.~\ref{subsec:beyond_qci}).}

% \begin{figure}[h]
%     \centering
%     \includegraphics[width=\linewidth]{figures/drift.pdf}
%     \caption{Hardware is known to have parameter drift for both single qubit gates (QB X) and for pairs of qubits (Edge Y) resulting in increasing error rates over time, though not necessarily monotonically. Errors on two qubit gates tend to be more volatile. On the left, we generate variations of known drift patterns \cite{} over arbitrary time steps to observe how expected program success rate declines over time (bottom). Calibration events (dashed black line) can restore the system to its former quality. In these cases, both circuit compilation and kernel compilation become stale meaning we must periodically restart the loop, forcing kernel re-compilation. Unfortunately, drift is not predictable and requires either full calibration or a more flexible routine like Optimus.}
%     \label{fig:drift}
% \end{figure}

\subsection{\revision{Hardware Drift}}
\revision{Device parameters drift over time, resulting in calibration information becoming stale and increasing error rates \cite{proctor2020detecting}. 
% Even if compiled carefully according to known error rates, tight control of system parameters is infeasible and therefore after enough time, these circuits will fail more frequently.  
Parameters are frequently updated after rounds of calibration. Both the circuit and the kernel should be recompiled to account for a potentially new device error landscape.}

% \revision{In Fig~\ref{fig:drift}, we demonstrate both example gate drift rates and its effect on EPS. If drift is not accounted for at all, systems quickly fall out of acceptable range and programs become virtually impossible to execute with any confidence. Forced re-calibration can correct for this, resetting error rate trajectories. 
\revision{Because drift is not uniform it can be unnecessarily expensive to re-calibrate the entire system, calibrating both parameters that have remained ``in-spec'' as well as those which have fallen ``out-of-spec.'' To accommodate this, more flexible and dynamic calibration schemes have emerged, such as Optimus \cite{kelly2018physical}, which constructs a calibration dependency graph which is periodically probed to determine which parameters to re-calibrate. Low shot-count overhead probes can be used to determine what has drifted out-of-spec. The size of this graph should scale quadratically with the number of qubits in a system with all-to-all connectivity.}

\begin{figure}[h]
    \centering
    \includegraphics[width=\linewidth]{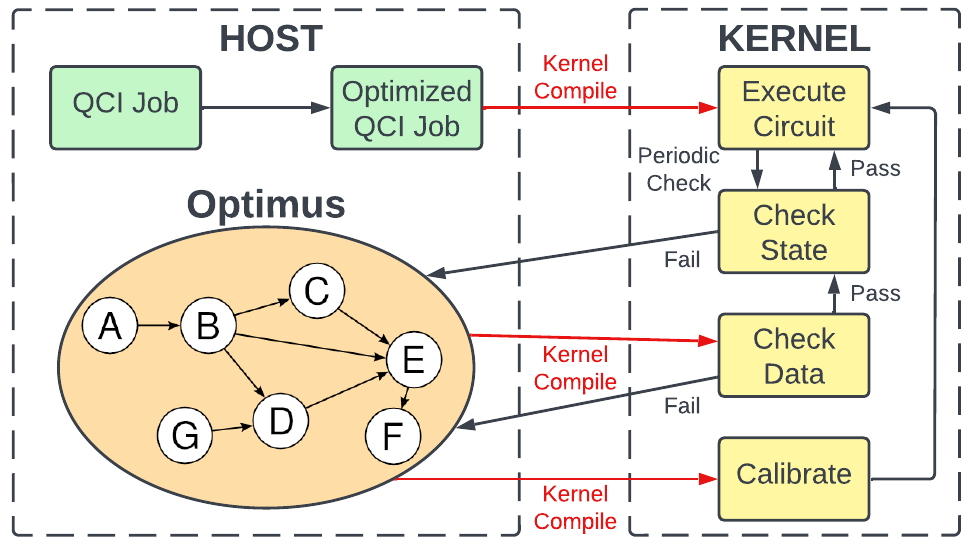}
    % \vspace{-1.5\baselineskip}
    \caption{\revision{Optimus is a flexible, graph-based re-calibration routine that periodically checks hardware parameters and re-calibrates those that have fallen ``out of spec.'' These routines require kernel compilation (indicated in red) to run the system probes (check data) and the calibration procedure apart from compiling the circuit kernel. When using the baseline approach each of these kernel compilations are repeated depending on the number of system probes, and size of the calibration graph in addition to multiple circuit runs. \ac{dlpc} reduces these kernel compilations for Optimus and circuit execution.}}
    % \vspace{-1\baselineskip}
    \label{fig:vqe_optimus_infra}
\end{figure}

\revision{
% , either in terms of hardware runtime overhead or compiling the kernels for
% if we exit the host-kernel internal loop to queue either 
% the light-weight probes or the calibration routine itself. 
Even when done quickly \cite{tornow2022minimum}, 10-15 experiments per node in the Optimus graph can lead to excessive kernel compilation overheads, increasing with the rate of drift which determines the frequency nodes will fail their probes. Following from \cite{riesebos2021universal}, we construct sparse random Optimus graphs with various failure rates for each node. We select random nodes as the starting point and follow the Optimus algorithm and use the expected calibration times reported in \cite{tornow2022minimum} for both the number of experiments (number of times kernel compilation is executed in the baseline case) and the time to complete these experiments. \ac{dlpc} avoids kernel re-compilation for calibration routines. We estimate probe execution time to be an order of magnitude cheaper, though the exact cost of these probes is not specified in either prior works \cite{riesebos2021universal, kelly2018physical}. The goal of this experiment is to demonstrate how drift can be both probed and corrected dynamically, however with potentially large overhead which is reduced by integrating DLPC. The accommodation of Optimus in \ac{dlpc} is summarized in Fig~\ref{fig:vqe_optimus_infra}.}

\begin{figure}[h]
    \centering
    \includegraphics[width=\linewidth]{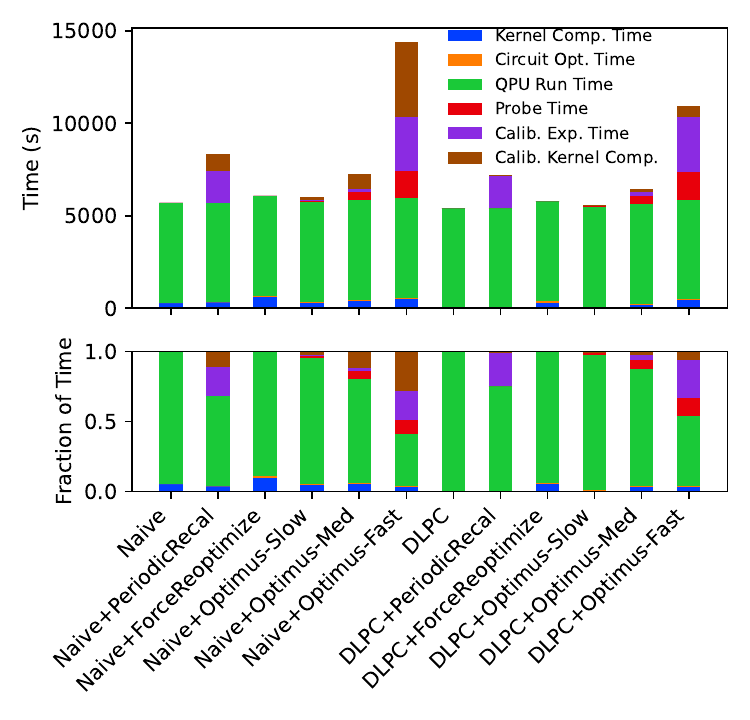}
    \vspace{-1.5\baselineskip}
    \caption{\revision{Execution time of activities when running under different calibration and optimization routines, both without (Naive) and with DLPC. Here VQE runs with hundreds of iterations each with 1000 shots. After each iteration, we randomly query an Optimus graph (as in \cite{riesebos2021universal}) which probabilistically reports failure relative to the drift rate. On failures, we run calibration of the node using experiment counts and execution times reported from \cite{tornow2022minimum}. As drift increases, both the time spent calibrating and the total number of experiments, hence kernel compilations, increases. With DLPC we remain in the kernel and accommodate dynamic re-calibration, reducing the total fraction (bottom) of time spent doing classical tasks.}}
    % \vspace{-1\baselineskip}
    \label{fig:vqe_optimus}
\end{figure}

\revision{In Fig~\ref{fig:vqe_optimus}, we compare various schemes including periodic forced recalibration of each routine in the Optimus graph (here it is infrequent, hence lower than the Optimus approach with fast drift), forced circuit optimization, and the generalized Optimus approach with random probes at various drift rates. Compared to the non-calibration versions of either naive or DLPC, these will necessarily be more expensive. However, notably DLPC has significant reductions in kernel compilation times, specifically due to the integrated probe and calibration routines. In this experiment, we explore a small qubit VQE instance (5 qubits) to solve for a single bond angle with several hundred iterations and with shot counts approximated by \cite{claudino2020benchmarking, gonthier2022measurements, gu_shot_budget} (order of thousands per iteration). Because of the random query model, we report the average over hundreds of samples. With DLPC, both the total execution time and the fraction of time spent compiling the kernel is substantially smaller compared to the baseline implementation. Because the underlying circuit is small, circuit reoptimization takes a very small fraction of the total runtime.}

\section{Related Work}
\label{sec:related}
\revision{Compilation of quantum programs has been studied extensively, though primarily at the circuit and pulse level. In \cite{gokhale_variational}, they explore the use of partial compilation at the pulse level by dividing the circuit structure into blocks which either are or are not re-compiled into pulses every iteration. Regardless of which block, the resulting pulses must still then be converted to basic waveforms and machine code to be played out by the control hardware.
% , i.e. they stop short of entirely flexible changes to machine code as the pulses change. 
Our work enables this, extending partial compilation to machine code.
% , and complements this prior work well. 

Circuit parameterization is not entirely new, for example, it is currently part of the OpenQASM 3.0~\cite{openqasm3} spec, however, these are entirely high-level representations, not executable instructions, and must be converted into hardware-specific executables. Work done at Rigetti \cite{Karalekas_2020} proposes parametric compilation at the \ac{isa} level. Our work proposes an execution model that relies on work presented by them, as \ac{dlpc} is only possible if the underlying architecture supports partially compiled parametric binaries. In contrast, work out of Delft \cite{krol2022efficient} proposes parametric compilation at the level of the QASM compiler in their stack, and does not extend it to the device level compilation. The quantum-classical interfacing problem has also been explored before, for example with IBM's Qiskit Runtime \cite{qiskit_web}, which enables users to submit entire jobs with interleaved classical processing rather than just circuits. This is primarily focused on reducing the queue time overheads between individual iteration executions for a single contiguous hardware allocation. However, this does not address the problem of kernel startup or compilation to classical control signals. Our work complements each of these prior works by addressing compilation lower in the pipeline. The batching work by Burch et al \cite{Burch_Sandia} addresses parametric compilation at the device level using pre-computation. They generate and compile a batch of programs using pre-computed parameters before running them contiguously. They are unable to perform batching and save on compilation costs if the parameters cannot be pre-computed, which our work can accommodate. Furthermore, we extend each of the above works by showcasing the adaption of parametric compilation beyond variational algorithms.}

\section{Conclusion}
\label{sec:conclusion}

While most current architectural work for quantum computers has focused on high-level machine and pulse optimizations, device-level code optimization has remained largely untouched meaning naive implementations of pulse-to-hardware instruction translation results in unnecessary compilation overhead. This work proposes a \ac{dlpc} technique that takes advantage of the shared structure and kernel-level invariants of quantum operations. This allows the device-level machine code to be compiled only once, with the changing parameters passed into the binary file through cheap \acp{rpc}. This technique results in the reduction of compilation overhead for the class of \ac{qci} problems.
% , as they have a parameterized quantum program that runs repeatedly with changing parameters. 
The technique can also be extended to iterative quantum programs that do not require any interleaved classical input but do have a shared program structure across runs.

We demonstrate this technique on a trapped-ion system and in simulation. We ran simple single-qubit \ac{vqe} programs on the hardware for a speedup of up to 2.7x using \ac{dlpc}. We demonstrated \ac{dlpc} on multi-qubit systems by running optimal pulse calibration routine in simulation and get a runtime speedup of up to 7.7x. We also demonstrate \ac{dlpc} for system characterization routine \ac{rb} and see a runtime speedup of up to 2.6x. \revision{We evaluate it on other platforms, quantum computing workflow in the cloud, and in cases requiring re-optimization and re-calibration.} Our proposed technique is most beneficial on a subset of \ac{qci} problems where a large number of circuits need to be run for a small number of shots. This is due to the growing domination of circuit execution time over compilation time as the number of shots increases. However, with faster gate times and \revision{improved} shot budget requirements, the advantage presented by \ac{dlpc} will grow.
\section*{Acknowledgment}
The work was funded by the National Science Foundation (NSF) STAQ Project (PHY-1818914, PHY-2325080), EPiQC - an NSF Expeditions in Computing (CCF-1832377), NSF Quantum Leap Challenge  Institute for Robust Quantum Simulation (OMA-2120757), the Office of the Director of National Intelligence, Intelligence Advanced Research Projects Activity through ARO Contract W911NF-16-1-0082, and the U.S. Department of Energy, Office of Advanced Scientific Computing Research QSCOUT program. Support is also acknowledged from the U.S. Department of Energy, Office of Science, National Quantum Information Science Research Centers, and Quantum Systems Accelerator.

%%%%%%%%% -- BIB STYLE AND FILE -- %%%%%%%%
% \clearpage
\bibliographystyle{IEEEtran}
\bibliography{main}
%%%%%%%%%%%%%%%%%%%%%%%%%%%%%%%%%%%%

\end{document}